\documentstyle[12pt]{article}

\begin{document}
\thispagestyle{empty}
{\hfill Preprint JINR E2-91-328}\vspace{2cm} \\
\begin{center}
\large\bf Nonlinear realizations of $W_3$ symmetry
\end{center}
\vspace{2cm}
\begin{center}
{\large E.Ivanov, S.Krivonos and A.Pichugin}\footnote{
Dniepropetrovsk State University, Ukraine}\vspace{1.8cm}\\
{\it Laboratory of Theoretical Physics,} \\
{\it Joint Institute for Nuclear Research, Dubna, Head Post Office}\\
{\it P.O. Box 79, 101 000 Moscow, Russia}
\end{center}
\vspace{1.5cm}
\begin{center}
\bf Abstract
\end{center}

We deduce the $sl_{3}$ Toda realization of classical $W_3$ symmetry
on two scalar fields in a geometric way, proceeding
from a nonlinear realization of some associate higher-spin symmetry
$W_{3}^{\infty}$. The Toda equations
are recognized as the constraints singling out a two-dimensional fully
geodesic subspace in the initial coset space of $W_{3}^{\infty}$.
The proposed geometric approach can be extended to other nonlinear
algebras and integrable systems.
\begin{center}
Submitted to {\it Phys. Lett. B}
\end{center}
\vfill
\begin{center}
Dubna 1991
\end{center}
\setcounter{page}0
\setcounter{footnote}0
\newpage

{\bf 1. Introduction.}  Recently, classical versions of Zamolodchikov's
$W_{N}$ algebras \cite{a1} received much attention (see, e.g.,
\cite{a12} - \cite{a5}). They were found to be the symmetry algebras of
some $2D$ field-theory models, such as the systems of free bosonic fields
\cite{{a2},{a3}} and Toda systems \cite{a4}, and certain steps towards
understanding
their geometric origin have been done \cite{a5}. In view of growing interest
to $W$-gravities, $W$-strings and related theories it is extremely
important to fully reveal, from various points of view, the geometries
underlying these $W$ symmetries.

In the present letter we suggest a new geometric
set-up for classical $W_{N}$ symmetries which relies upon the
nonlinear realization approach \cite{{a6},{a7}} and is a generalization of
the
treatment of $W_{2}$ (Virasoro) symmetry in ref. \cite{a8}. We address
here the simplest nontrivial case $N=3$,
however our construction seems equally applicable to
other algebras and superalgebras of this kind. The basic trick
allowing us to apply the standard nonlinear realization
scheme to the algebras of the type $W_{N}$ consists in
replacing them by some infinite-dimensional linear
$W_{\infty}$ type algebras ($W_{N}^{\infty}$ in what follows)
which arise if one treats as independent all the composite higher-spin
generators appearing in the commutators of the basic $W_{N}$ generators.
We deduce the $sl_{3}$ Toda
realization of $W_{3}$ \cite{a3} in a purely geometric way as a particular
coset realization of $W_{3}^{\infty}$ and show that the $sl_{3}$ Toda
equations are also intimately related to the intrinsic geometry of
$W_{3}^{\infty}$: they single out a two-dimensional fully geodesic surface
in the coset space of $W^{\infty}_{3}$. \\

{\bf 2. From $W_{3}$ to $W_{3}^{\infty}$.}  The most general classical
$W_{3}$ algebra is defined by the following relations \cite{a12}
\begin{eqnarray}
\left[ L_n, L_m \right] & = & (n-m)L_{n+m} +
\frac{c}{12}(n^3-n)\delta_{n+m,0} \nonumber \\
\left[ L_n, J_m \right] & = & (2n-m)J_{n+m} \nonumber \\
\left[ J_n, J_m \right] & = & - 24 (n-m)\Lambda_{n+m}-
   \frac{n-m}{2}\left[ (n+m)^2-\frac{5}{2}nm-
4 \right]L_{n+m} - \nonumber \\
        & & -\frac{c}{48}(n^2-4)(n^2-1)n\delta_{n+m,0} \quad , \label{27}
\end{eqnarray}
where $L_{n},\;J_{n}$ are the spin 2 (Virasoro) and spin 3 generators and
\begin{equation}
\Lambda_n = \frac{1}{c}\sum_mL_{n-m}L_m \; .
\end{equation}

The nonlinear realization techniques
we intend to apply to this $W_{3}$ symmetry have been worked out in
\cite{a6} for symmetries based on Lie algebras, i.e. linear
algebras. In order to generalize these techniques to
the $W_{N}$ type algebras we propose to treat the spin 4 generators
$\Lambda_{n}$ in (\ref{27}) and all other
higher-spin composite generators present in the enveloping algebra
of $L_{n}, J_{m}$ as independent ones. In other words, one replaces
(\ref{27}) by some linear infinite-dimensional higher-spin algebra
$W_{3}^{\infty}$
\begin{equation}
W_{3}^{\infty} = \left\{ L_{n},\; J_{n}, \; \Lambda_{n},\;...\;,\;J^{S}_{n},
\;... \right\}, \;\;\;\;S = 5,6, ...
\end{equation}
in which the commutation relations between generators of the lowest
spins (2 and 3) are given by (\ref{27}) and all the remaining relations
involving the higher-spin generators $\{ \Lambda_{n}, \ldots ,
J^{S}_{n}, \ldots \}$ are successively evaluated proceeding
from these basic relations and the quadratic relation (2).
In principle, any commutator can be specified in this way and the
generic form of higher-order commutators can be indicated\footnote{It would
be interesting to compare $W_{3}^{\infty}$ (and $W^{\infty}_{N}$) with the
$W_{\infty}$ algebras of ref. \cite{{b0},{a10}}. It is likely that they belong
to different
classes. For example, $W_{3}^{\infty}$ involves several independent types of
the spin 6 generators, while in $W_{\infty}$ each spin occurs once.}.
For our
purpose it is of no need to know the detailed structure of these
commutators. It will be crucial that $c$ in (1) is nonvanishing and composite
generators form a closed set (the last statement follows from a direct
inspection of all possible products of the basic spin 2 and spin 3 currents).

Let us list some important subalgebras of (3). Of major relevance for our
purpose
is the following centerless ifinite-dimensional subalgebra which is the
genuine generalization of the truncated Virasoro algebra
$\left\{ L_{-1}, \;L_{0}, \; L_{1}, \ldots \right\}$
\begin{equation}
\begin{array}{rcl}
       L_{-1}& L_{0} & L_{1}\quad L_{2} \quad \ldots \\
   J_{-2}\quad J_{-1}& J_{0} & J_{1} \quad J_{2} \quad \ldots \\
 \Lambda_{-3}\quad   \Lambda_{-2}\quad \Lambda_{-1}& \Lambda_{0} &
\Lambda_{1} \quad \Lambda_{2} \quad \ldots \\
 \ldots\ldots\ldots\ldots \ldots\ldots \ldots& \ldots &
\ldots \ldots \ldots \ldots \ldots \quad .
\end{array}
\end{equation}
We call it ${\widetilde W}_{3}^{\infty}$. To see that (4) is a subalgebra,
one represents (1) and (3) as the algebra of generalized variations
generated by an element $\sum_S \int dz a^{S}(z)\;J^{S}(z), \;
J^{S}(z) = \sum_nz^{-n-S}J^{S}_{n},\;S=2,3,...$ and notices that (4)
is singled out by restriction to the parameters-functions
$a^{S}(z)$ regular at $z=0$. Then closure of (4) follows from the fact
that the variety of such functions is closed under differentiation and
multiplication. To avoid a possible confusing, let us point out that the
higher-spin
generators in (4), when treated as composite, still involve the products of
generators $L_{n},\; J_{m}$ with all negative and positive conformal
dimensions.

{}From the fact that (3) contains also a subalgebra consisting of the
generators
which are obtained from those in (4) by the reflection $n \rightarrow -n$
one immediately finds out the existence
of the wedge type \cite{{b0},{b1}} subalgebra
$W_{\wedge}$ in
$\widetilde{W}_{3}^{\infty}$
\begin{equation}
W_{\wedge}= \left\{
\begin{array}{c}
L_{-1} \quad L_{0} \quad L_{1} \\
J_{-2} \quad J_{-1} \quad J_{0} \quad J_{1} \quad J_{2}  \\
\Lambda_{-3} \quad \Lambda_{-2} \quad \Lambda_{-1} \quad \Lambda_{0}
 \quad \Lambda_{1} \quad \Lambda_{2} \quad \Lambda_{3} \\
\ldots\ldots \ldots \ldots\ldots\ldots \ldots \ldots  \ldots \ldots \ldots \\
\end{array}   \right\}
\end{equation}
(dots mean higher-spin generators with proper indices).
One can check (using (1), (2) and the property that the
composite generators form a subalgebra) that all generators in (5) with spin
$\geq 4$
constitute
an infinite-dimensional ideal and the factor-algebra of (5) over this ideal is
$sl(3,R)$
\begin{equation}\label{210}
W_{\wedge} /\{ \Lambda_{-3},\ldots ,\Lambda_{3},\ldots,J^{S}_{-S+1}\ldots
J^{S}_{S-1},\ldots  \}
  \sim   sl(3,R) \quad .
\end{equation}

{\bf 3. Nonlinear realizations of $W_{3}^{\infty}$.}  Having replaced
$W_{3}$ by a linear algebra $W_{3}^{\infty}$, we are
ready to construct a nonlinear realization of the
$W_{3}^{\infty}$ symmetry, following the
generic prescriptions of \cite{{a6},{a7}}.
By reasonings of simplicity and for the correspondence with the
consideration in ref. \cite{a8} we limit ourselves
to the truncated algebra $\widetilde{W}_{3}^{\infty}$ (4).
Our ultimate aim will be to show that the $sl_3$
Toda realization of $W_{3}$ on two $2D$ scalar fields \cite{a3} and the Toda
equations of motion themselves result
from a particular coset realization of $\widetilde{W}_{3}^{\infty}$ after
imposing certain covariant constraints on the relevant coset
parameters.

In order to have manifest $2D$ Lorentz
symmetry, we start from the product  of two commuting copies of
the $\widetilde{W}_{3}^{\infty}$ symmetry groups (we denote it $G$)
with the algebra
\begin{equation}
{\cal G} = \widetilde{W}_{3+}^{\infty} \times \widetilde{W}_{3-}^{\infty}
\quad .
\end{equation}
As a next step we need to choose an appropriate coset of $G$,
which is actually reduced
to specifying the stability subgroup $H$. We have checked that
no finite-dimensional cosets
exist in $G$. So,
in the present case one deals with infinite-dimensional coset manifolds.

Fortunately, it is
not so difficult as it could seem. First, by reasonings of the reducibility
to the Virasoro case \cite{a8} the coset should include the generators
$\{ L^{\pm}_{-1},L^{+}_{0} + L^{-}_{0}, L^{\pm}_{1}, ...L^{\pm}_{n},\}$,
with the $2D$ Minkowski space coordinates $x^{\pm}$
and the
Liouville field $u(x)$
being parameters corresponding to the first three generators\footnote{We
treat all the coset
parameters other than $x^{\pm}$ as fields given on $x^{\pm}$; in other words,
we actually deal
with a two-dimensional hypersurface in the coset space. This does not
influence the transformation properties of coset parameters and is typical
for nonlinear realizations of space-time symmetries \cite{a6}.}. Secondly,
having as a goal to eventually come to the $sl_{3}$ Toda theory, we need
to reserve a place for the second Toda field $\phi(x)$ as a coset
parameter. The only
appropriate generators are $J^{\pm}_{0}$,
so we are led to include a linear combination of them into the set of the
coset generators. Finally, it would be desirable to place
all higher-spins generators into the stability subgroup and
in what follows not to care about them.

These three requirements are satisfied in a minimal way with
the two-parameter family of stability subgroups generated by
\begin{eqnarray}
{\cal H} & = & \{ J_{-2}^{\pm}-\alpha \beta \;J_{2}^{\mp}\;,\;\;
J_{-1}^{+}+\frac{\sqrt 3}{2}L_{-1}^{+}-\alpha(J_1^{-}+
\frac{\sqrt{3}}{2}L_1^{-})\;,\;\;
J_{0}^{+}-J_{0}^{-}\;,\;\; L_0^+ - L_0^-\;,  \nonumber \\
& & J_{-1}^{-}-\frac{\sqrt 3}{2}L_{-1}^{-}
 -\beta(J_1^{+}-
\frac{\sqrt{3}}{2}L_1^{+});\;
 \Lambda_{-3}^{\pm}\;,\;\Lambda_{-2}^{\pm}\;,\ldots ; \ldots ;
J^{S\pm}_{-S+1}\;, \ldots \}\;,\; S \geq 5. \label{31}
\end{eqnarray}
The combinations of the generators $L$ and $J$ in (8) turn out to form
the Borel subalgebra of the diagonal $sl(3, R)$ in the sum of two commuting
factor-algebras $sl(3,R)$ (6) contained in $\widetilde{W}_{3+}^{\infty}$
and $\widetilde{W}_{3-}^{\infty}$. The parameters $\alpha , \;\beta $ reflect a
freedom in extracting this diagonal $sl(3,R)$.

Now, an element of the coset space $G/H$ can be parametrized as follows
\begin{equation}\label{32}
g\equiv G/H=e^{x^{\pm}L^{\pm}_{-1}}e^{\psi^{\pm}_{1}J^{\pm}_{1}}
 e^{\xi^{\pm}_{1}L^{\pm}_{1}}e^{\psi^{\pm}_{2}J^{\pm}_{2}}\ldots\;
e^{u(L^{+}_{0}+L^{-}_{0})}e^{\phi (J^{+}_{0}+J^{-}_{0})} \quad .
\end{equation}
Here,  $u(x), \phi(x), \psi^{\pm}_{1}(x),\xi^{\pm}_{1}(x),\ldots $ constitute
an infinite tower of the coset parameters-fields. The group $G$ acts on the
coset (\ref{32}) from the left
\begin{equation}\label{33a}
g_{0}(\lambda )g(x,u,\phi,\ldots )=g(x',u',\phi',\ldots )\cdot h \quad ,
\end{equation}
where $g_{0}(\lambda )$ is an arbitrary element of $G$, and $h$ belongs to the
subgroup $H$. The arrangement of the group factors in (\ref{32}) is
convenient in that the transformation laws of $2D$ coordinates $x^{\pm}$
and the fields $u(x),\;\phi (x)$ under conformal
transformations ($g_{0}=\exp\sum_{n=-1}^{+\infty}\lambda_nL_n$)
are of the standard form
\begin{eqnarray}
\delta_{\lambda} x^{\pm} & = & \lambda^{\pm}\left( x^{\pm} \right) =
   \sum_{n=-1}^{+\infty}\lambda^{\pm}_n\left( x^{\pm} \right)^{n+1} \nonumber\\
\delta_{\lambda} u(x) & = & u'(x')-u(x)=\frac{1}{2}(\partial_{+}\lambda^{+}+
   \partial_{-}\lambda^{-} ) \nonumber \\
\delta_{\lambda} \phi(x) & = & \phi'(x')-\phi(x)= 0 \quad . \label{35}
\end{eqnarray}

The $J_{n}$ transformations of the coordinates
and parameters-fields can be also deduced from the general formula (10).
For $x^{\pm}$ and $u(x),\;\phi(x)$ we get the following transformations
(we write down here only the
transformations generated by the ``+'' branch of the group $G$)
\begin{eqnarray}
\delta_{a} x &=& -\frac{\sqrt 3}{2}\; a'(x) +2{\sqrt 3}\;
(\xi_{1} +
\frac{\sqrt 3}{2}   \psi_{1})\;a(x)
\nonumber \\
\delta_{a} u(x) & = &
-{{\sqrt 3}\over 2}\;(\xi_1 +
\frac{\sqrt 3}{2}\psi_1)\;a'(x)+
  ( 6\psi_2 +2{\sqrt 3}\xi_1^2 )\;a(x) \nonumber \\
\delta_{a}\phi (x) & = & {1\over 4}a''(x)-
{3\over 2}\;(\xi_{1} + \frac{\sqrt 3}{2}\psi_1)
\;a'(x)-3\;[\; \xi_{2} - (\xi_{1} + \frac{\sqrt 3}{2}\psi_1)^2\;]\;
a(x) \quad , \label{36}
\end{eqnarray}
where the function $a(x)$ collects constant parameters of
the group element $g_0$
$$
g_0=\exp {\sum_{n=-2}^{+\infty}a_nJ_n} \quad , \quad
a(x) = \sum_{n=-2}^{+\infty}a_n x^{n+2} \quad .
$$
and, for brevity, we omitted the index ``+''of $x$ and higher-spin coset
fields. Note that for non-zero $\alpha\;,\;\beta$ in (\ref{31}) these
transformations
turn out to have a nontrivial action on the
coset fields belonging to the ``--'' light-cone branch of $G$. For instance,
the fields $\xi^{-}_1,\;\psi^{-}_1$ transform as
\begin{eqnarray} \label{36a}
\delta \psi^{-}_1 = \frac{\sqrt{3}}{2}\;\delta \xi^{-}_1 =
{1\over 2}\;\alpha (a'- 4 a \xi^{+}_{1})\;
e^{-2(u + \sqrt{3}\phi )}
\end{eqnarray}
(for $J$-transformations from the ``--'' branch the situation is reversed,
with $\alpha$ replaced by $\beta$).

The main peculiarity of transformations (\ref{36}), (\ref{36a}) and
their crucial difference from the conformal ones is that $J_n$ have no
realizations on the coordinates $x^{\pm}$ alone -- these generators
necessarily mix the coordinates with the coset fields
$\psi_1,\psi_2,\xi_1,\xi_2$. Moreover, in fact we deal here with an
infinite-dimensional nonlinear representation of $\widetilde{W}_{3}^{\infty}$
because the fields $\psi_1,\xi_1$ are
transformed through higher-spin fields $\psi_2,\xi_2,\psi_3,\xi_3$ and so on.
We stress that at this step the active form of
transformations of the coset fields contains in each term
no more than
one derivative on fields, the latter arising due to the field-dependent
shift of $x$
in (12). The same is true, of course, for higher-spin transformations
which are generated via Lie brackets of (12), (13).

The fundamental geometric objects of nonlinear realizations, covariant
Cartan's forms, are introduced by the standard relation
\cite{a6}
\begin{equation}\label{37}
g^{-1}dg=\sum_{n=-1}^{+\infty}\omega^n_{\pm}L_n^{\pm}+
   \sum_{n=-2}^{+\infty}\Omega^n_{\pm}J_n^{\pm}+ \ldots,
\end{equation}
where dots stand for the forms entering with the higher-spin generators.
For our further purposes it will be important
that the infinite set of forms associated with the coset generators is closed
under the left shifts (\ref{33a}). Let us explicitly give a few first forms
\begin{eqnarray}
\omega_{-1}^{\pm} & = & e^{-u} ch({\sqrt 3}\phi )dx^{\pm}
\quad , \quad \omega_{0}^{\pm} = du - 2\xi_{1}^{\pm}dx^{\pm}
\nonumber \\
\Omega_{-2}^{\pm}  & = & 0 \quad , \quad
\Omega_{-1}^{\pm}   =  -\frac{2}{\sqrt 3}e^{-u}
sh({\sqrt 3}\phi )dx^{\pm}  \quad , \quad
\Omega_{0}^{\pm}    = d\phi -3\psi_{1}^{\pm}dx^{\pm}\;.
\label{39}
\end{eqnarray}
We see that the fields $\psi^{\pm}_{1},\;\xi^{\pm}_{1}$ enter into
some coset space Cartan
forms linearly and homogeneously. A straightforward inspection shows that
this is a general phenomenon: for each coset field except for
$u(x),\;\phi(x)$ one can indicate such a Cartan form. Their set is
\begin{equation}
\omega_0^{+}+\omega_0^{-}\;, \;\;
\Omega_0^{+}+\Omega_0^{-}\;,\;\;
\omega_n^{+}|_{+} \;, \;\; \omega_n^{-}|_{-} \; , \;\;
\Omega_n^{+}|_{+} \;, \;\;   \Omega_n^{-}|_{-}
\quad \mbox{for all }n\geq 1 \quad , \label{310}
\end{equation}
where the symbols $|_{\pm}$ label the projections of a given form onto
$dx^{\pm}$, respectively. \\

{\bf 4. $sl_{3}$ Toda from $W^{\infty}_{3}$}. So far our coset fields carry
no any dynamics, their origin is purely geometric: they define the
embedding of $2D$ Minkowski space $\{x^{\pm}\}$ as a hypersurface
into an infinite-dimensional coset space of the group
$G=\widetilde{W}^{\infty}_{3+} \times \widetilde{W}^{\infty}_{3-}$.
The latter generates motions of this
two-dimensional
hypersurface in the coset space. Now we wish to show that after imposing
an infinite number of $G$-covariant constraints on the coset fields this
hypersurface is completely specified by the two fields, $u(x)$ and $\phi(x)$,
which are subject to the $sl_{3}$ Toda equations and for which
$G$-transformations take the standard form of the $W_{3+}\times W_{3-}$
ones \cite{a3}.

What we are going to effect is the covariant reduction procedure
worked out in \cite{a8} and applied there for constructing a coset
realization of $W_{2}$ symmetry on a single (Liouville) field $u(x)$. In our
case it goes as follows. Given Cartan
forms (\ref{37}) defined from the
beginning on the whole algebra
${\cal G} = \widetilde{W}^{\infty}_{3+} \times \widetilde{W}^{\infty}_{3-}$,
one constrains them as
\begin{equation}\label{41}
g^{-1}dg=\sum_{n=-1}^{+\infty}\omega^n_{\pm}L_n^{\pm}+
   \sum_{n=-2}^{+\infty}\Omega^n_{\pm}J_n^{\pm}+ \ldots
= g^{-1}_{red}dg_{red}\in
   \widetilde{\cal G} \quad ,
\end{equation}
where $\widetilde{\cal G}$ is some subalgebra containing the stability
algebra ${\cal H}$
defined in eq. (\ref{31}). In the $W_{2}$ case \cite{a8} $\widetilde{\cal G}$
was chosen to be $sl(2,R)$ and this immediately led to the Liouville equation
for $u(x)$. In the case at hand it is natural to choose
\begin{eqnarray}
\widetilde{\cal G} & = & \{
R^{\pm}=J^{\pm}_{-1}+\frac{\sqrt 3}{2}L^{\pm}_{-1}-\alpha
   ( J^{\mp}_1+\frac{\sqrt 3}{2}L_1^{\mp})\;,\;
S^{\pm}=J^{\pm}_{-1}-\frac{\sqrt 3}{2}L^{\pm}_{-1}-\beta
   ( J^{\mp}_1-\frac{\sqrt 3}{2}L_1^{\mp}) \nonumber \\
&& B^{\pm} = J^{\pm}_{-2}-\alpha\beta\; J_2^{\mp}, \;
U=L_0^{+}-L_0^{-},\;
T=J_0^{+}-J_0^{-},\;
 \mbox{All higher-spin generators} \}. \label{42}
\end{eqnarray}
This algebra is an extension of ${\cal H}$ by two generators, $R^{-}$
and $S^{+}$. It is
easy to see (remembering the reasonings of Sec.2) that all the higher-spin
generators in (\ref{42}) form an ideal and that
the factor-algebra of (\ref{42}) over this ideal is the diagonal
$sl(3,R)$ in the sum of the two commuting factor-algebras $sl(3,R)$ (6)
(coming
from $\widetilde{W}_{3+}^{\infty}$ and $\widetilde{W}_{3-}^{\infty}$). So
this choice
is indeed a natural generalization of the option of ref. \cite{a8}.

The covariant reduction condition (\ref{41}) amounts to an infinite set of the
constraints
\begin{eqnarray}
\omega_1^{\pm} & = & -{1\over 2}(\alpha + \beta)\omega_{-1}^{\mp}-
               {\sqrt{3} \over 4}(\alpha - \beta)\Omega_{-1}^{\mp}\;,\;\;
\Omega_1^{\pm}  =
-{1\over 2}(\alpha + \beta)\Omega_{-1}^{\mp} -
{1\over \sqrt{3}}(\alpha - \beta)\omega_{-1}^{\mp}
 \nonumber \\
\Omega_2^{\pm} & = & \alpha \beta\; \Omega_{-2}^{\mp}\;, \;\;\;\;
\omega^{+}_{0} + \omega^{-}_{0} = 0\;,\;\;\;\; \Omega^{+}_{0} +
\Omega^{-}_{0} = 0 \nonumber \\
\omega_n^{\pm}  &=&  0 \;, \;\;\;\;\; \Omega_{n+1}^{\pm}  =  0\;,\;\;\;\;\;
 \mbox{for all }n\geq 2 \quad . \label{43}
\end{eqnarray}
Eq. (\ref{41}) and its detailed form (\ref{43}) are covariant under
left shifts (\ref{33a}), which can be explicitly checked with making
use of the relations (1). This guarantees that the original group structure
is not destroyed.

One observes that each of eqs. (\ref{43}) gives
rise to two equations, for the projections of the
$\omega,\;\Omega$ onto $dx^{+}$ and $dx^{-}$. An important subset of these
equations is formed by the equations implying the projections (\ref{310})
to vanish. They are purely
algebraic and serve to express higher-spin coset fields in terms of
$u(x),\;\phi(x)$ and derivatives of the latter
(``inverse Higgs effect'', ref. \cite{a7}). All such coset fields can
be eliminated in this manner, e.g.,
\begin{eqnarray}
\xi_1^{\pm}& = & \partial_{\pm}u(x) \quad , \quad
\xi_2^{\pm} =  \frac{1}{3}\left[ \partial_{\pm}^2u(x)+
    (\partial_{\pm}u(x))^2+
(\partial_{\pm}\phi (x))^2 \right]\nonumber \\
\psi_1^{\pm} & = & \frac{2}{3}\partial_{\pm}\phi (x) \quad , \quad
\psi_2^{\pm} =  \frac{1}{6}\partial_{\pm}^2\phi (x)+
     \partial_{\pm}u(x)\partial_{\pm}\phi (x) , \quad \mbox{etc.}\label{311}
\end{eqnarray}

The remaining equations prove to be dynamical: they constrain $u(x)$, $\phi(x)$
to obey the $sl_{3}$ Toda equations:\footnote{We are at liberty
to put $\alpha$ and/or $\beta$ equal to zero, which corresponds to some
contractions of $\widetilde{\cal G}$ (18). This gives rise to
alternative reductions, with $u+\sqrt{3}\phi$ and/or $u-\sqrt{3}\phi$
subject to free equations.}
\begin{eqnarray}
\partial_{+}\partial_{-}u & = &
-{1\over 2}\alpha e^{-2(u+\sqrt{3}\phi )}  -
 {1\over 2}\beta e^{-2(u-\sqrt{3}\phi )} \nonumber \\
\partial_{+}\partial_{-}\phi & = &
-{\sqrt{3}\over 2}\alpha e^{-2(u+\sqrt{3}\phi )}  +
 {\sqrt{3}\over 2}\beta e^{-2(u-\sqrt{3}\phi )} \quad .
\label{eq}
\end{eqnarray}

Now we substitute the expressions (\ref{311}) into the geometric
$\widetilde{W}^{\infty}_{3}$ transformation laws (\ref{36}) and find the
resulting spin 3 transformations of the fields $u(x),\;\phi (x)$
\begin{eqnarray}
\tilde\delta u(x)\equiv u'(x)-u(x) & = &
-\frac{1}{2}a'(x)\partial\phi +
a(x)\left( \partial^2\phi+4\partial u\partial\phi\right)   \nonumber \\
\tilde\delta \phi (x) \equiv \phi'(x)-\phi (x)& = &
\frac{1}{4}a''(x)-
\frac{3}{2}a'(x)\partial u +a(x)\left(2(\partial u)^2 -2(\partial\phi )^2-
\partial^2 u \right) \quad . \label{312}
\end{eqnarray}
Together with the spin 2 transformations (\ref{35}) they constitute the
standard $sl_{3}$ Toda realization of classical $W_{3}$ symmetry (1), (2)
\cite{a3} (to be more precise, its regular part with the parameters-functions
regular at $x=0$). The appropriate currents
\begin{eqnarray}
\gamma^{-2}T^{(-2)}& = & -\frac{1}{2}(\partial u)^2-\frac{1}{2}
(\partial\phi )^2- \frac{1}{2}\partial^2 u \nonumber \\
\gamma^{-2}J^{(-3)} & = & \frac{1}{4}\partial^3\phi
+\frac{3}{2}\partial^2\phi\partial u
   +\frac{1}{2}(\partial u)^2\partial\phi +2(\partial u )^2\partial\phi-
   \frac{2}{3}(\partial\phi )^3  \label{33}
\end{eqnarray}
generate, via the Poisson brackets, the whole algebra (1), (2) with
$c = 3\gamma^{2} $.

Thus we have shown that the Toda realization of nonlinear
$W_{3}$ symmetry can be reproduced
proceeding from a pure
geometric coset realization of some
associate linear higher-spin symmetry $\widetilde{W}_{3}^{\infty}$.
In other words, $W_{3}$ can be viewed as
a particular field realization of this infinite-dimensional linear algebra.

In contrast to the $W_2$ (Liouville) case \cite{a8}, the algebraic and
dynamical
constraints in (19) are in general mixed under $G$ transformations, so
the exact $W_{3+}\times W_{3-}$ structure comes out only on shell, when
eqs. (21) are fulfilled. The $sl_3$ Toda action is still invariant
under the transformations (22) with the parameters-functions of both
light-cone chiralities, and the related Lie bracket structure generates
two independent algebras $W^{\infty}_3$. However, these algebras do not
commute with each other; the commutativity is restored only on shell. It
is interesting to seek for another form of the $sl_3$ Toda action which
would possess the $W^{\infty}_{3+}\times W^{\infty}_{3-}$ symmetry off shell
and originally involve an infinite number of auxiliary fields in parallel
with $u(x)$ and $\phi (x)$. The whole set of constraints (19) could be
then expected to follow from this action as the field equations.

It is worth noting that the zero-curvature representation for eqs.
(\ref{eq}) on the $sl(3,R)$ algebra \cite{a13} automatically arises in our
approach.
Indeed, after
imposing the constraints (\ref{41}), (\ref{43}), we are left with the form
valued in the algebra
$\widetilde{\cal G}$ (\ref{42}) and involving only the fields
$u(x),\;\phi (x)$:
\begin{equation}
\Omega_{red} =
\Omega_{sl(3,R)} + \ldots \quad . \label{45}
\end{equation}
Here
\begin{eqnarray}
\Omega_{sl(3,R)}&=& \frac{1}{\sqrt 3}\;e^{-u-{\sqrt 3}\phi}\;dx^{\pm}
 R^{\pm}-\frac{1}{\sqrt 3}\;e^{-u+{\sqrt 3}\phi}\;dx^{\pm}S^{\pm}+\nonumber\\
& & +(\partial_{-}udx^{-}-\partial_{+}udx^{+})U
 +(\partial_{-}\phi dx^{-}-\partial_{+}\phi dx^{+})T
\label{45a}
\end{eqnarray}
and dots stand for the terms with higher-spin generators.
As we have already mentioned, the higher-spin
generators
constitute an ideal in $\widetilde{\cal G}$, so
the Maurer-Cartan equation for the form $\Omega_{sl(3,R)}$
closes without any reference to the higher-spin components.
Thus, the Maurer-Cartan
equation for $\Omega_{red}$ immediately results in the zero-curvature
condition for $\Omega_{sl(3,R)}$:
\begin{equation}\label{46}
d^{\mbox{ext}}\Omega_{sl(3,R)}=\Omega_{sl(3,R)}\wedge\Omega_{sl(3,R)}\quad .
\end{equation}
It is a simple exercise to verify that eq.(\ref{46}) is equivalent to the
$sl_{3}$ Toda equations (\ref{eq}).

Finally, let us comment on the geometric meaning of the covariant reduction
procedure. As was explained in \cite{{a8},{a14}}, the
essence of this procedure consists in reducing a given
group space to its some lower-dimensional fully geodesic
subspace, in a way covariant with respect to the original nonlinear
realization.
The dynamical equations (Liouville equation \cite{a8} and others \cite{a14})
come out as the constraints
accomplishing this reduction (in parallel with the inverse Higgs
algebraic constraints). In the present case the relevant fully
geodesic subspace is the two-dimensional coset space $SL(3,R)\ /H$ (modulo
higher-spin generators) which is
thus the genuine analog of the pseudosphere $SL(2,R)\ /SO(1,1)$ figuring
in the $W_{2}$ example of ref. \cite{a8}.\\

{\bf 5. Conclusion.}
In this paper we have explicitly demonstrated that $W_3$ symmetry of
the $sl_3$ Toda system and field equations of the latter
have a remarkable geometric origin: they stem from a coset realization of
infinite-dimensional symmetry associated with the {\it linear} algebra
$W^{\infty}_3$. Doubtless, other Toda systems (based both on the algebras
$sl(N,R)$ and the algebras from other Cartan's series, e.g. $so(N)$) and
their $W_N$ symmetries admit an analogous interpretation in terms of the
appropriate $W_N^{\infty}$ algebras. The proposed approach could provide a
new
insight into the classical and quantum structure of the Toda-type theories
and, hopefully, lead to understanding all nonlinear algebras (and, perhaps,
quantum algebras), as well as the integrable hierarchies associated with them,
from a common geometric point of view. In the nearest perspective we are
planning to apply our construction to other nonlinear algebras and
superalgebras, e.g., Knizhnik-Bershadsky superalgebras \cite{a18}. On this
path we expect to obtain new integrable systems and to shed
more light on the geometric structure of the known ones, such as the KdV,
MKdV and KP hierarchies. \\

{\bf Acknowledgements.} It is a pleasure for us to thank S.Bellucci,
A.Isaev, J.Lukierski, V.Ogievetsky, G.Sotkov and
M.Vasiliev for interest in the work and discussions. E.I. \& S.K.
are grateful to the Theory Division of LNF-INFN in Frascati for hospitality
extended to them during the course of this work.

\end{document}